Stabilization and heteroepitaxial growth of metastable tetragonal FeS thin films by pulsed laser deposition

Kota Hanzawa[1)], Masato Sasase[2)], Hidenori Hiramatsu[1), 2), a)], and Hideo Hosono[1), 2)]

[1)] Laboratory for Materials and Structures, Institute of Innovative Research, Tokyo Institute of Technology, Japan
[2)] Materials Research Center for Element Strategy, Tokyo Institute of Technology, Japan
[a)] email: h-hirama@mces.titech.ac.jp

## Abstract

Pulsed laser deposition, a non-equilibrium thin-film growth technique, was used to stabilize metastable tetragonal iron sulfide (FeS), the bulk state of which is known as a superconductor with a critical temperature of 4 K. Comprehensive experiments revealed four important factors to stabilize tetragonal FeS epitaxial thin films: (i) an optimum growth temperature of 300 °C followed by thermal quenching, (ii) an optimum growth rate of ~7 nm/min, (iii) use of a high-purity bulk target, and (iv) use of a single-crystal substrate with small in-plane lattice mismatch ($CaF_2$). Electrical resistivity measurements indicated that none of all the films exhibited superconductivity. Although an electric double-layer transistor structure was fabricated using the tetragonal FeS epitaxial film as a channel layer to achieve high-density carrier doping, no phase transition was observed. Possible reasons for the lack of superconductivity include lattice strain, off-stoichiometry of the film, electrochemical etching by the ionic liquid under gate bias, and surface degradation during device fabrication.

## 1. Introduction

Exploration of new superconducting materials has been intensively reactivated in the superconductor research community since the discovery of an Fe-based superconductor, F-doped LaFeAsO, in 2008 [1] because of its high critical temperature ($T_c$) and unique pairing mechanism [2−4]. Through subsequent worldwide explorations of other Fe-based and related superconductors [5], numerous new materials have been discovered, and the maximum $T_c$ of the Fe-based family has reached 55 K [6], which is the second highest $T_c$ at ambient pressure among all superconductors; cuprates possess the highest $T_c$.

Among the Fe-based superconductor family, iron chalcogenides (Fe*Ch*) display some unique characteristics [7−9]. Fe*Ch* have the simplest chemical formula and crystal structure (Fig. 1(a)) consisting of only Fe and *Ch* with an anti-PbO-type tetragonal structure. The Fe*Ch* layer is composed of edge-sharing Fe*Ch*$_4$ tetrahedra; i.e. Fe*Ch* has no insertion layer between the Fe*Ch* layers, different from other Fe-based superconductors. The most studied Fe*Ch* is FeSe ($T_c$ = ~8 K in the bulk state [10]) because its $T_c$ can be greatly enhanced via various approaches such as structure variation and carrier doping. As examples of its structure variation, the $T_c$ of FeSe can increase to 36.7 K by applying external pressure [11] and 11.4 K through lattice strain introduced from single-crystal substrates via thin-film growth processes [12]. As for carrier doping, we demonstrated that the $T_c$ of FeSe was raised to 35 K by extremely high-density electron accumulation using electric-double layer transistor (EDLT) structures [13, 14]. This kind of carrier-induced high $T_c$ has also been reported in [15−18] and observed for monolayer-thick FeSe [9, 19].

Although tetragonal FeS (*t*-FeS) has the same crystal structure as that of FeSe, which is shown in Fig. 1(a), it was previously believed that *t*-FeS was not a superconductor but exhibited metallic or semiconductor-like behavior [20, 21]. However, *t*-FeS exhibiting superconductivity with $T_c$ = 4 K was reported recently [22]. Thus, it is expected that phenomena observed for FeSe, such as drastic enhancement of $T_c$, may also appear for *t*-FeS. To examine this possibility, an epitaxial thin film is the most promising platform because it is possible to introduce both tensile and compressive stresses as well as to controllably dope the film with a high density of carriers using EDLT structures.

It should be noted that *t*-FeS cannot be synthesized by conventional techniques such as a simple solid-state reaction between elemental Fe and S, from which only hexagonal phases (*h*-FeS; see Fig. 1(b)) are generated because they are thermodynamically much more stable than *t*-FeS in the Fe–S binary phase diagram [23]. Only unconventional methods such as deintercalation of alkali metal (*A*) from $A_x\text{Fe}_{2-y}\text{S}_2$ [24] are effective to obtain $t$-FeS. They indicate that $t$-FeS is a thermodynamically metastable phase.

Therefore, in this study, we use the non-equilibrium process, pulsed laser deposition (PLD), to stabilize $t$-FeS. Even though epitaxial FeSe$_{1-x}$S$_x$ solid solution films [25], polycrystalline [26], and monolayer-thick $t$-FeS films [27] have been fabricated, a pure $t$-FeS epitaxial thin film has not been reported. We presume that stabilization of $t$-FeS is quite sensitive to the thermal environment because it is a metastable phase, even though thermal assistance is necessary for crystallization and epitaxial growth of precursors. In this paper, by considering the influences of thermal, kinetic, and reaction paths, we succeed in stabilization of $t$-FeS and its epitaxial growth on a CaF$_2$ substrate. This is the first demonstration of stabilized pure $t$-FeS epitaxial thin films. We then perform high-density electron doping of the $t$-FeS thin films by constructing an EDLT structure.

## 2. Experimental procedure

➢ **Synthesis of polycrystalline bulk targets**

Polycrystalline bulk samples with nominal chemical compositions of FeS$_{1+x}$ ($x$ = 0, 0.1, 0.2, and 0.4) were synthesized for use as PLD targets via a solid-state reaction between elemental Fe (99.9%) and S (99.9999%). In an Ar-filled glove box, Fe and S powders were mixed with the appropriate mass ratio, sealed in evacuated silica-glass tubes, and then heated at 900 °C for 24 h. Then, the preheated Fe + S powders were pressed into pellets with a diameter of 7 mm and thickness of ~6 mm, sealed in evacuated silica-glass tubes, and heated at 900 °C for 36 h. The roughly estimated bulk density of the $x = 0$ sample was ~80%.

- **Thin film growth**

FeS thin films with a thickness of ~30 nm were grown by PLD in a vacuum growth chamber with a base pressure of $<5 \times 10^{-5}$ Pa. The film thickness for each PLD growth experiment was precisely determined based on the result obtained by x-ray reflectivity analysis. In this study we judged by some preliminary experiments that thinner thicknesses than 50 nm are appropriate for stabilization of the $t$-FeS phase because thicker films started to become the $h$-FeS phase. A KrF excimer laser with a wavelength of 248 nm and repetition rate of 10 Hz was used as an ablation source. The growth rate ($r_g$) was controlled between 3.3 and 7.5 nm/min by varying the laser fluence. Six kinds of (001)-oriented single crystals with dimensions of $10 \times 10 \times 0.5$ mm were used as substrates: Y-stabilized $ZrO_2$ (YSZ), $CaF_2$, $(LaAlO_3)_{0.3}(Sr_2AlTaO_6)_{0.7}$ (LSAT), $SrTiO_3$ (STO), GaAs, and MgO. YSZ, LSAT, and MgO were thermally annealed at 1350, 800, and 1100 °C in air, respectively, before use. STO was annealed at 1050 °C after wet etching with buffered HF solution [28] before use, whereas $CaF_2$ and GaAs were used without any preliminary treatment (i.e. used as-received from suppliers). During film growth, the substrates were heated at a substrate temperature ($T_s$) between room temperature (RT) and 400 °C from the back side of a substrate carrier made of Inconel alloy with a halogen lamp. The substrate heating was stopped just after film deposition, and then the films were rapidly cooled in the growth chamber.

> **Analysis of structure and chemical composition**

Crystalline phases and structures of polycrystalline bulk targets and obtained films were evaluated by X-ray diffraction (XRD) using Cu K$\alpha$ radiation. The amounts of $h$-FeS and impurity phases in the bulk samples were estimated by Rietveld analysis. A parallel-beam x-ray, which was monochromated by a two-bounce Ge (220) crystal, was used to measure rocking curves and asymmetric $\phi$ scans. The thickness of the films was precisely determined by x-ray reflectivity analysis using the parallel-beam geometry. Chemical compositions of the films were quantified from wavelength-dispersive x-ray fluorescence (XRF) analysis. The chemical composition of the film on $CaF_2$ was precisely determined by wavelength-dispersive electron probe microanalysis. The surface morphology of the films was observed by tapping-mode atomic force microscopy. The cross-sectional microstructure of a $t$-FeS film on $CaF_2$ was observed with a scanning transmission electron microscope (STEM) using high-angle annular dark field (HAADF) mode. During observation, we also performed selected-area electron diffraction (SAED) measurements in the $t$-FeS film regions and point composition analysis around the interface between $t$-FeS and $CaF_2$ by energy-dispersive x-ray (EDX) spectroscopy.

> **Electronic transport properties**

Electrical resistivities ($\rho$) of the $t$-FeS films were measured by a four-probe technique. Au films deposited with a direct-current sputtering system were used as contact electrodes. The measurement temperature ($T$) was varied from 2 to 300 K. For carrier doping of the $t$-FeS phase, we constructed an EDLT structure using a *t*-FeS film on $CaF_2$ as a channel layer. Gate voltages ($V_G$) of 0 to +5 V were applied through a Pt electrode at 220 K. Details of the measurement procedure and device configuration are the same as those reported in [13]. Transfer curves (i.e. drain current $I_D$ and gate leak current $I_G$ at a constant drain voltage $V_D$ of +0.1 V as a function of $V_G$) at 220 K and the dependence of the sheet resistance ($R_s$) of the channel on $T$ in the range of 2–220 K were measured for the EDLT structure.

## 3. Results and discussion

To stabilize metastable *t*-FeS thin films, we used PLD as a non-equilibrium growth process. Through experimental trials, we found three critical factors affecting the formation of *t*-FeS—temperature, kinetics, and chemical composition of PLD targets.

First, because *t*-FeS is a metastable phase, we carefully investigated the influence of temperature on the grown crystalline phases. Figure 2(a) depicts the dependence of crystalline phases grown using a stoichiometric FeS target (i.e. nominal $x = 0$) on $T_s$. At high $T_s$ (400 °C), only thermally stable *h*-FeS phases (space group: $P\bar{6}2c$ (No. 190) and/or $P6_3/mmc$ (No. 194)) were detected; i.e. *t*-FeS was not generated. In contrast, at $T_s$ between RT and 300 °C, *t*-FeS formed with preferential orientation along the *c*-axis for the out-of-plane direction (i.e. diffractions perpendicular to the film plane). This result indicates that $T_s$ needs to be lower than 400 °C for stabilization of $t$-FeS. We then thermally annealed a $t$-FeS film grown at $T_s$ = 300 °C in a vacuum growth chamber for 3 h immediately after finishing the deposition. Figure 2(b) compares the XRD pattern of the as-grown $t$-FeS film with the annealed one. Although only the $t$-FeS phase was observed before annealing, $h$-FeS phases appeared after annealing. This structural phase transition induced by thermal annealing has also been reported for a $t$-FeS powder [29], indicating that thermal quenching just after deposition is also necessary to protect the generated $t$-FeS phase after deposition and prevent conversion to stable $h$-FeS phases.

Next, we examined effect of $r_g$ at $T_s$ = 300 °C on the obtained crystalline phases because kinetics is also another general factor to overcome thermal equilibrium and stabilize metastable phases. Figure 2(c) shows the films grown at different $r_g$. When $r_g$ was too low (3.3 nm/min) or too high (7.5 nm/min), $h$-FeS phases emerged. In contrast, the pure $t$-FeS phase was observed at moderate $r_g$ (6.8 nm/min). These results indicate that the optimum $r_g$ is ~7 nm/min, which is another critical parameter to effectively nucleate and stabilize $t$-FeS. Therefore, we tentatively concluded that the optimum stabilization conditions of $t$-FeS are $T_s$ = ~300 °C followed by quenching and $r_g$ = ~7 nm/min. However, subsequent XRF analysis indicated that the obtained $t$-FeS contained large amounts of excess Fe; i.e. the [Fe]/[S] atomic ratio was ~1.1.

According to the above XRF result, we then tried to tune the [Fe]/[S] ratio to reach stoichiometry by adding excess S to the PLD bulk targets. The nominal compositions of PLD targets we examined were $FeS_{1+x}$ ($x$ = 0, 0.1, 0.2, and 0.4). Other parameters were all fixed; i.e. $T_s$ = ~300 °C followed by quenching and $r_g$ = ~7 nm/min. Figure 3(a) presents XRD patterns of $FeS_{1.2}$ (top) and stoichiometric FeS (bottom) bulk targets. In the case of the stoichiometric FeS target, $h$-FeS (space group: $P\bar{6}2c$) and a small amount of Fe (volume fractions of 98.4 and 1.6 wt%, respectively) were observed. The $FeS_{1.2}$ target is composed of three phases, $Fe_7S_8$ (70.6 wt%), $h$-FeS (22.5 wt%), and $FeS_2$ (6.9 wt%). Figure 3(b) shows the XRD patterns of FeS films grown using the four different targets with nominal $x$ = 0–0.4. While the stoichiometric FeS target led to formation of pure $t$-FeS, all the other S-rich targets did not generate $t$-FeS at all; only $h$-FeS. This result indicates that the chemical composition of the films cannot be easily controlled by adding large amounts of excess S to the PLD target. In addition, the crystalline phase of the target is also probably very important for the formation of $t$-FeS as the phase transition from phases such as $FeS_2$ and $Fe_7S_8$ to $t$-FeS would be more difficult than that from $h$- to $t$-FeS because of the higher activation energy of the former transitions. Thus, we concluded that the most important factors to stabilize the $t$-FeS phase grown via PLD are $T_s$ = 300 °C followed by thermal quenching, $r_g$ = ~7 nm/min, and the chemical composition of bulk target should be pure $h$-FeS.

    Even though the $t$-FeS phase was successfully stabilized on STO and YSZ single-crystal substrates under the optimum growth conditions, the in-plane orientation was random (i.e. preferentially $c$-axis oriented only in the out-of-plane direction); i.e. the $t$-FeS films were not epitaxial. We therefore attempted $t$-FeS deposition on various kinds of single-crystal

substrates with different in-plane lattice mismatches under the optimum growth conditions in an attempt to achieve epitaxial *t*-FeS film growth. Figure 4(a) shows the XRD patterns of films grown on six kinds of substrates. On YSZ, CaF$_2$, LSAT, and STO, single-phase *t*-FeS was obtained; whereas on GaAs and MgO, impurity *h*-FeS phases were segregated. We then examined the in-plane orientation of the films. Figures 4(b) and (c) display the results of $\phi$-scans for the *t*-FeS 112 diffractions of the films on STO and LSAT, respectively. In the case of STO, no diffractions were detected, indicating that *t*-FeS grew with preferred orientation only for the *c*-axis (out-of-plane); i.e. random for in-plane. This feature was observed also for the films grown on GaAs and YSZ substrates (data not shown). Conversely, in the case of LSAT, we observed diffractions consistent with the four-fold symmetry of the tetragonal lattice, which can be classified into two groups rotated by 37° with respect to each other. The rotation angle of 37° is relatively close to the 45° usually observed for tetragonal lattice. The exact origin of such angle difference from 45° is unclear at present. However, we show a possible origin of the unusual rotated domain by 37° in the supplementary information (see supplementary information available at [URL]); whereas this consideration is just a speculation and one of the possibilities only from the view point of their atomic configurations. This result indicates that there are two single-crystalline domains in the film. Figure 4(d) shows the $\phi$-scans for *t*-FeS 111 and CaF$_2$ 111 diffractions. Because four-fold symmetry originating from a single in-plane domain rotated by 45° with respect to $CaF_2$ was observed, we concluded that $t$-FeS grew heteroepitaxially on $CaF_2$. The epitaxial relationship is [001] $t$-FeS ∥ [001] $CaF_2$ for the out-of-plane direction and [100] $t$-FeS ∥ [110] $CaF_2$ for the in-plane direction. This is the first demonstration of epitaxial $t$-FeS thin-film growth. Therefore, another important factor is added to the critical conditions to stabilize epitaxial $t$-FeS thin films; i.e. use of a $CaF_2$ (001) single crystal as a substrate.

To examine the reason for the observed selective epitaxial growth of $t$-FeS on $CaF_2$, the structure parameters of the systems are summarized in table 1. Here, we mainly focused on the in-plane lattice mismatch ($\Delta a$) between $t$-FeS ($a_{FeS}$ = 3.683 Å [24]) and the substrate ($a_{sub}$) calculated from $\Delta a = (a_{sub} - a_{FeS})/a_{sub}$. When $\Delta a \geq 7.61\%$, the in-plane symmetry was random. With decreasing $\Delta a$, domains with long-range order appeared. Epitaxial growth was realized at $\Delta a < \sim 6.72\%$. For the system with YSZ, $a_{sub}$ had two kinds of lattice parameters; primitive (5.139 Å) and $1/\sqrt{2}$ times the lattice parameter (3.638 Å). In the case of the latter multiple lattice, the absolute value of $\Delta a$ is the smallest in table 1 if we expect heteroepitaxial growth with 45° rotation. However, despite the small $\Delta a$ of YSZ, we did not observe any in-plane orientation of the $t$-FeS film grown on YSZ. Therefore, we speculate that the large $\Delta a$ = 28.4% for cube-on-cube heteroepitaxy affects the growth of $t$-FeS, resulting in random in-plane orientation.

Although we used substrates with larger $a_{sub}$ values than that of $a_{FeS}$, the $c$-axis lattice parameter of all the obtained $t$-FeS films was expanded (e.g. $c$ = 5.092 Å on CaF$_2$) compared with that of a $t$-FeS single crystal (5.034 Å [24]), leading to $t$-FeS films with compressed in-plane lattice parameters. The $a$-axis lattice parameters of the films (we can estimate them only for the films on LSAT and CaF$_2$) were small ($a$ = 3.668 Å on CaF$_2$) compared with that of a $t$-FeS single crystal (3.683 Å [24]). The above strain trend for out-of-plane and in-plane directions is completely opposite to that of ~10 nm-thick $t$-FeSe on STO [30]. We are unsure of the exact reason of such unusual lattice variation; however, the misoriented in-plane lattice, excluding $t$-FeS on CaF$_2$, and an interface layer (see that will be displayed in Fig. 5) may be possible origins. The chemical composition of all the FeS films was off-stoichiometric; i.e. the [Fe]/[S] atomic ratio was > 1.0 except for that on CaF$_2$ ([Fe]/[S] = 0.9). To evaluate the crystallinity of the $t$-FeS films, we measured the full width at half maximum of rocking curves for the out-of-plane $t$-FeS 001 diffraction ($\Delta\omega$; see table 1). $\Delta\omega$ of $t$-FeS on CaF$_2$ (2.00°) was the best of the systems, which was attributed to its epitaxial growth. Figures 4(e)–(k) show the surface morphologies of the $t$-FeS thin films. Although droplets derived from PLD were observed in the films grown on LSAT and GaAs, all the surfaces were relatively flat; e.g. the root mean square roughness ($R_{rms}$) of $t$-FeS on CaF$_2$ was 1.6 nm.

Next, we confirmed the heteroepitaxial growth of the $t$-FeS film on CaF$_2$ also by STEM measurements in HAADF mode. Figure 5(a) is a wide-view cross-sectional HAADF-STEM image of the system. Clear stacking along the $c$-axis, which is consistent with the layered crystal structure of *t*-FeS (Fig. 1(a)), was observed in the entire film region from the substrate–film interface toward film surface. Diffraction spots of the SAED pattern in the film region (inset of Fig. 5(a)) are assigned to the *t*-FeS phase, which supports heteroepitaxial growth of the *t*-FeS film and is consistent with the XRD results (Fig. 4). Furthermore, the existence of a 3–3.5 nm thick interface diffusion layer between the *t*-FeS film and $CaF_2$ was unveiled (Fig. 5(b)). Similar interface layers have also been observed between *t*-Fe(Se, Te) and $CaF_2$ [31, 32]. We then performed point analysis by EDX to roughly estimate the chemical composition of the film (Fig. 5(b)). The results suggested that the *t*-FeS film bulk region was almost stoichiometric and all of the constituent elements diffused in the interface layer. However, we were unable to determine the exact structure and composition of this interface layer. Figure 5(c) shows an atomic-resolution image in the *t*-FeS film region. All bright positions were assigned to Fe (orange) and S (yellow) in *t*-FeS. Additionally, dark line-shaped areas were also alternately stacked along the *c*-axis, which is consistent with the fact that *t*-Fe*Ch* has no insertion layer between the edge-sharing Fe*Ch*$_4$ tetrahedra layers (Fig. 1(a)). According to the results of microstructure observation, we concluded that the film grown on $CaF_2$ is *t*-FeS with respect to not only its averaged structure (i.e. XRD results) but also to its local atomic coordination structure.

    The electronic transport properties of the films were then examined. Figure 6 shows the dependence of $\rho$ of the *t*-FeS films grown under the optimum conditions on five kinds of single-crystal substrates on *T*. Even

though it was reported that *t*-FeS bulk exhibits superconductivity at 4 K [22], none of the thin films exhibited superconductivity down to 2 K; instead, the films displayed insulator-like behavior as previously suggested [20] and as observed for very thin strained FeSe films [30, 33, 34]. Absolute $\rho$ values of the films strongly depended on substrate type over the whole $T$ range. We speculate that one of the possible origins of this behavior is the crystallinity of the films because the $\rho$ values of samples on GaAs, STO, and YSZ are very high compared with those on the other substrates. According to the results in table 1, these three samples exhibit only out-of-plane *c*-axis orientation, mainly because of their large $\Delta a$. Conversely, the crystallinity of samples on LSAT and $CaF_2$, which also exhibited lower $\rho$, was much higher than that of films on the other substrates. Here it should be noted that the structure of all the *t*-FeS films obtained was expanded along the *c*-axis and contracted along the *a*-axis compared with the corresponding values for a *t*-FeS single crystal. In a *t*-FeSe thin film, in-plane lattice variation should affect superconductivity more strongly than out-of-plane lattice variation [35]. Moreover, superconductivity in *t*-FeS bulk disappears by applying external pressure [36]. Thus, in-plane compressive strain is a possible origin of the lack of superconductivity and insulator-like behavior of the *t*-FeS films. Another possibility is their off-stoichiometric chemical composition [22]. Therefore, if we can release lattice strain and/or optimally tune chemical composition, our *t*-FeS films may display superconductivity.

    Finally, we attempted high-density carrier doping of a *t*-FeS film using an EDLT structure. We used $t$-FeS grown heteroepitaxially on $CaF_2$ as a channel layer because superconductivity induced by EDLT strongly depends on the quality of the channel layer, as we previously reported for a $t$-FeSe EDLT structure [14]. Figure 7(a) presents the transfer curves of the EDLT structure under applied $V_G$ of up to +5.0 V at 220 K. We induced clear modulation of $I_D$ that was three orders of magnitude larger than that of $I_G$, even though the on/off ratio was small (~5%). In the cyclic measurements, large hysteresis loops were observed. Furthermore, the initial $I_D$ at $V_G$ = 0 V was not the same as that at the end point and decreased from the first cycle to the second, indicating that the resistance of the $t$-FeS channel increased during the cyclic measurements. The phase transition of $t$-FeS was then examined by measuring $R_s$ under $V_G$. Figure 7(b) shows the dependence of $R_s$ on $T$ at $V_G$ of 0 to +5 V. Even under a high $V_G$ of +5.0 V, no phase transition was induced. With increasing $V_G$, $R_s$ also increased, which indicates that carriers were not doped in the channel. However, we observed $I_D$ modulation in the transfer curves (Fig. 7(a)). Therefore, carriers are doped in the channel but the channel thickness and/or surface state changed during the measurements. Possible reasons for the changes in the channel during cycling may be electrochemical etching by the ionic liquid [15] and surface degradation, similar to the case for $t$-FeSe [14, 37]. Thus, further improvement of film quality and the fabrication process of the EDLT is necessary to induce superconductivity in $t$-FeS EDLT structures.

## 4. Conclusion

We stabilized metastable $t$-FeS via the thin-film growth process PLD. The essential factors to stabilize the metastable $t$-FeS phase were determined and included using an optimum growth temperature of $T_s$ = 300 °C followed by thermal quenching, optimum $r_g$ of ~7 nm/min, and pure $h$-FeS bulk target. At high $T_s$, competitive $h$-FeS, which is the thermodynamically stable phase, preferentially nucleated, whereas at $T_s$ < 400 °C, the $t$-FeS phase was stabilized. Because $t$-FeS is metastable, thermal quenching immediately after deposition was necessary to stabilize the phase and prevent segregation of competitive $h$-FeS. The optimum $r_g$ was ~7 nm/min because $h$-FeS was mainly observed at other $r_g$. Moreover, it was clarified that using a high-purity $h$-FeS ($P\bar{6}2c$) PLD target was important to obtain $t$-FeS. When other PLD targets mainly composed of $Fe_7S_8$ were used, growth of the $t$-FeS phase was not stabilized. These results suggest that growth of $t$-FeS is very sensitive to growth conditions. We then revealed that epitaxial growth of $t$-FeS was achieved only on a $CaF_2$ single-crystal substrate, which was probably related to the small in-plane lattice mismatch between $t$-FeS and $CaF_2$. Therefore, another important factor for $t$-FeS epitaxial growth is selection of a single-crystal substrate with small in-plane lattice mismatch. This report presented the first demonstration of epitaxial $t$-FeS thin-film growth. None of the fabricated $t$-FeS thin films exhibited superconductivity even though superconductivity at 4 K was previously reported for bulk $t$-FeS. The lack of superconductivity was attributed to the introduced compressive in-plane and tensile out-of-plane lattice strain, and/or off-stoichiometric chemical composition. In addition, no phase transition was induced even in an EDLT structure with $t$-FeS grown on $CaF_2$ as a channel layer, which was possibly because of electrochemical etching by the ionic liquid and/or surface degradation during device fabrication. Perspectives to induce superconductivity in the strained $t$-FeS epitaxial films that we expect could be that an opposite direction strain (i.e. tensile one along in-plane and compressive one along out-of-plane) and/or almost completely relaxed growth could be effective by employing appropriate buffer layer between $t$-FeS and $CaF_2$ and/or performing appropriate initial treatments just before growth such as thermal annealing and plasma treatment.

## Acknowledgments

This work was supported by the Ministry of Education, Culture, Sports, Science, and Technology (MEXT) through the Element Strategy Initiative to Form Core Research Center. H. Hi. was also supported by the Japan Society for the Promotion of Science (JSPS) through Grants-in-Aid for Scientific Research (A) and (B) (Grant Nos. 17H01318 and 18H01700), and Support for Tokyotech Advanced Research (STAR).

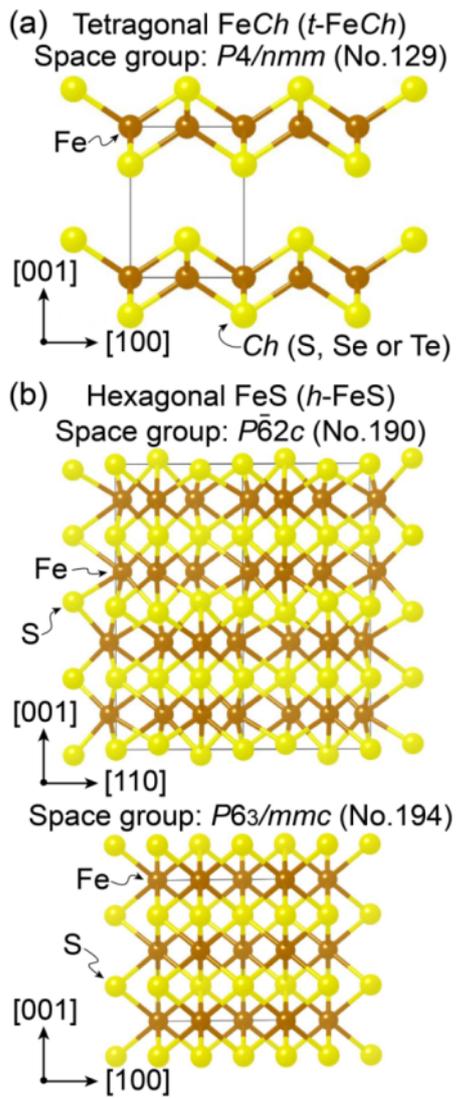

**Figure 1.** Crystal structures of FeS. (a) Tetragonal phase (*t*-Fe*Ch*, *Ch* = S, Se, or Te). (b) Two types of hexagonal phases (*h*-FeS).

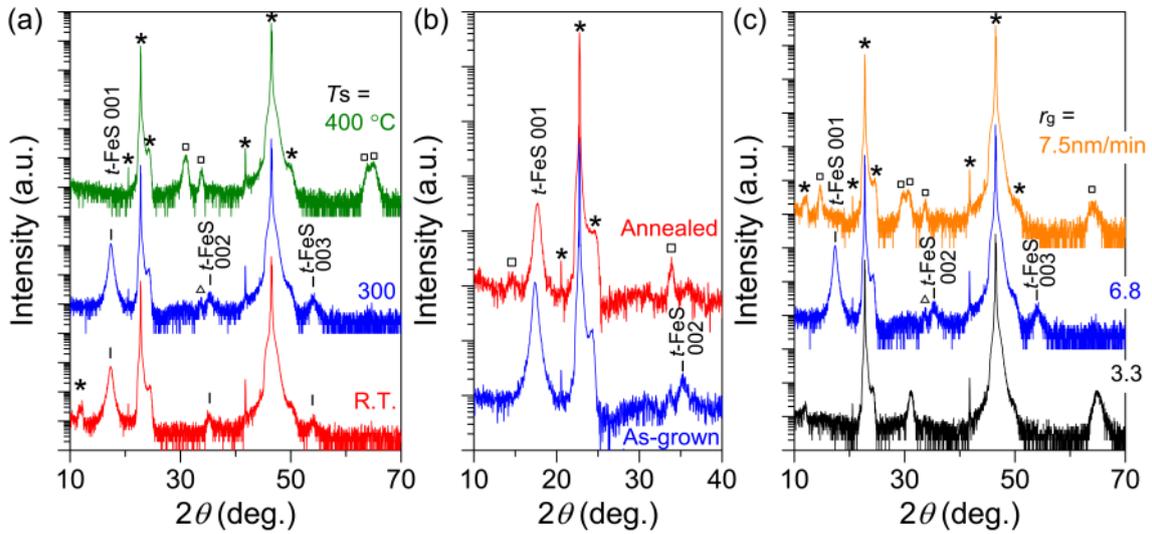

**Figure 2.** Crystalline phase analysis based on out-of-plane XRD patterns of FeS films grown on STO substrates under different conditions. Vertical bars, asterisks, and black squares denote diffraction positions of $t$-FeS, substrate, and $h$-FeS, respectively. Possible origins of a weak diffraction peak at $2\theta = $ ~33.6°, indicated by open triangles, are $10\bar{1}1$ of $h$-FeS ($P6_3/mmc$, No. 194), $11\bar{0}2$ of $h$-FeS ($P\bar{6}2c$, No. 190), and/or 200 of cubic $FeS_2$ ($Pa\bar{3}$, No. 205). (a) Films grown at $T_s$ of RT, 300, and 400 °C. (b) Influence of post-deposition thermal annealing under vacuum at 300 °C for 3 h on the phase of a $t$-FeS film. Bottom and top patterns are for as-grown and annealed films, respectively. (c) Films grown at $T_s$ = 300 °C and $r_g$ = 3.3, 6.8, and 7.5 nm/min.

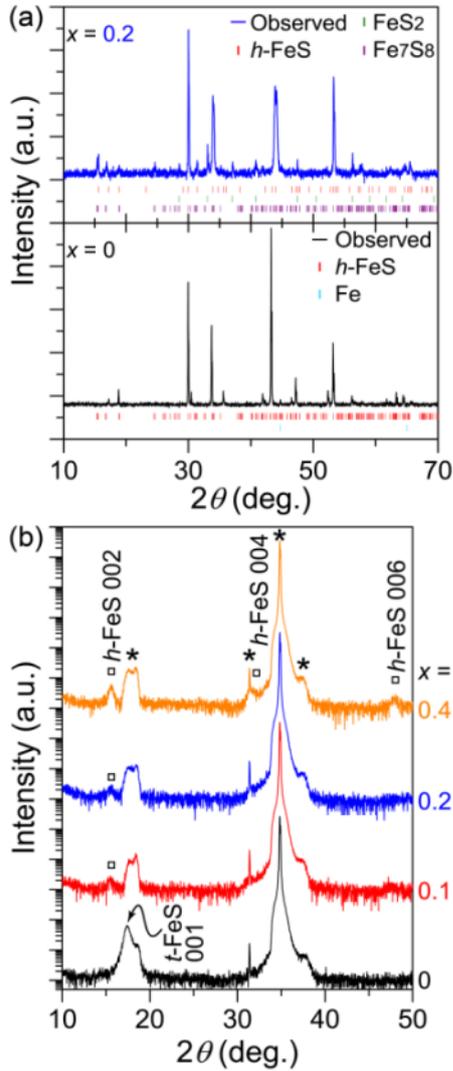

**Figure 3.** Influence of the chemical composition of PLD bulk targets on the formation of crystalline phases via growth at $T_s$ = 300 °C and = ~7 nm/min. (a) XRD patterns of polycrystalline targets with nominal chemical compositions of $FeS_{1.2}$ (top) and FeS (bottom). (b) Out-of-plane XRD patterns of FeS films on YSZ grown using four kinds of PLD targets with nominal chemical compositions of $FeS_{1+x}$ ($x$ = 0, 0.1, 0.2, and 0.4). The arrow at $2\theta$ = 17.6° indicates the 001 diffraction of the $t$-FeS phase. Asterisks and black squares denote diffraction positions of the substrate and $h$-FeS, respectively.

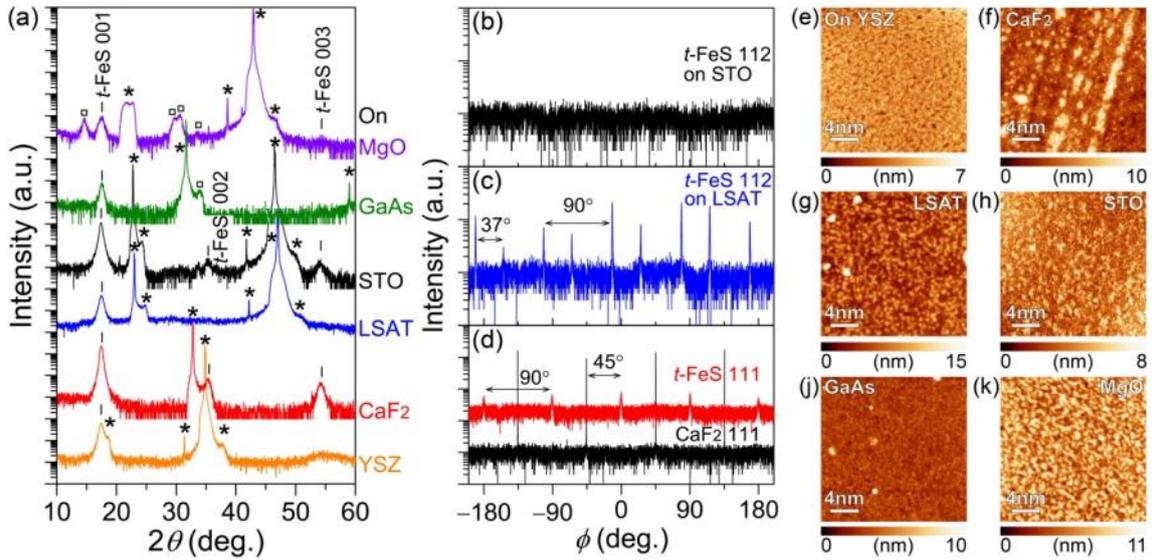

**Figure 4.** Influence of substrate type on FeS thin-film growth. (a) Out-of-plane XRD patterns of the films grown on different substrates. Vertical bars, asterisks, and black squares denote diffraction positions of *t*-FeS, substrate, and *h*-FeS, respectively. In-plane symmetry of *t*-FeS, evaluated by $\phi$-scans, grown on (b) STO, (c) LSAT, and (d) CaF$_2$. Surface morphologies of FeS films deposited on (e) YSZ, (f) CaF$_2$, (g) LSAT, (h) STO, (i) GaAs, and (j) MgO. Horizontal color bars are height scales.

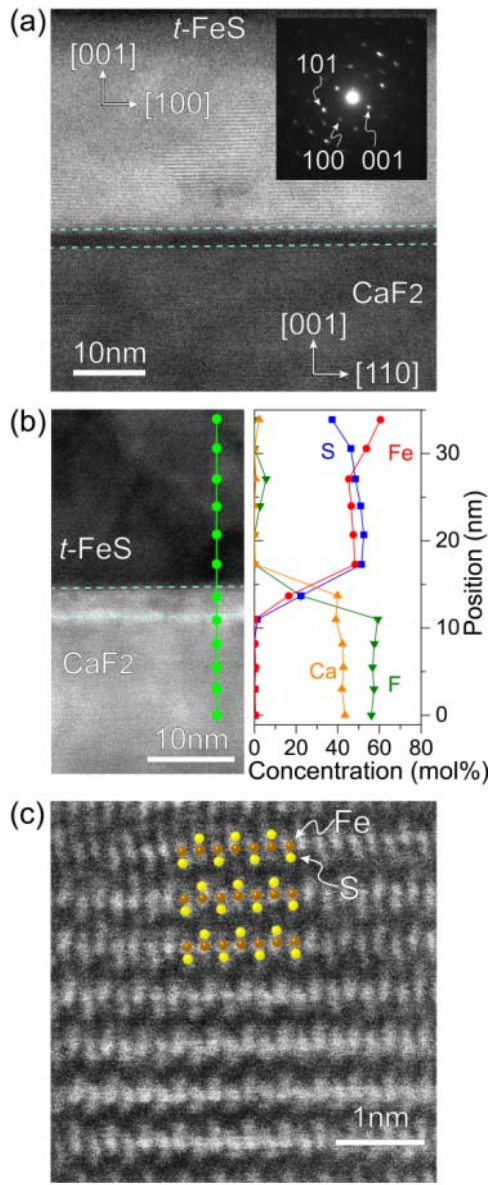

**Figure 5.** Cross-sectional microstructure and chemical composition of a *t*-FeS film on $CaF_2$ measured by HAADF-STEM. The direction of the incident electron beam was along [010] of the *t*-FeS film. (a) Wide-view cross-sectional HAADF-STEM image. Light blue dashed lines denote the boundaries between *t*-FeS or $CaF_2$ and an interface diffusion layer. Inset is the SAED pattern taken from the *t*-FeS film area. (b) EDX point analysis (right) at the positions shown by light green circles in the left image. (c)

Atomic-resolution image of the *t*-FeS film region. Orange and yellow spheres indicate the positions of Fe and S atoms in *t*-FeS, respectively.

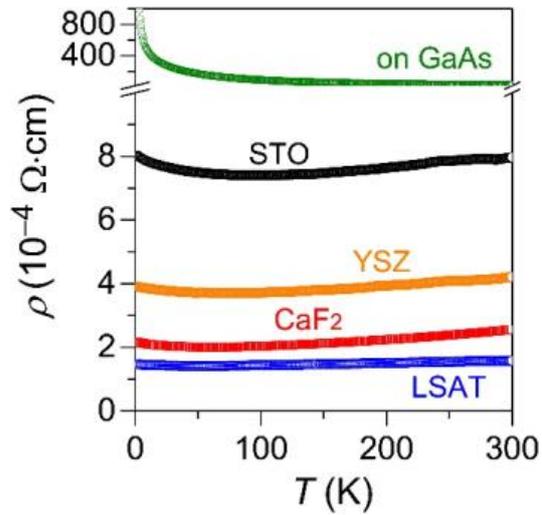

**Figure 6.** $\rho$–$T$ curves of the *t*-FeS films grown under the optimum conditions on five kinds of single-crystal substrates.

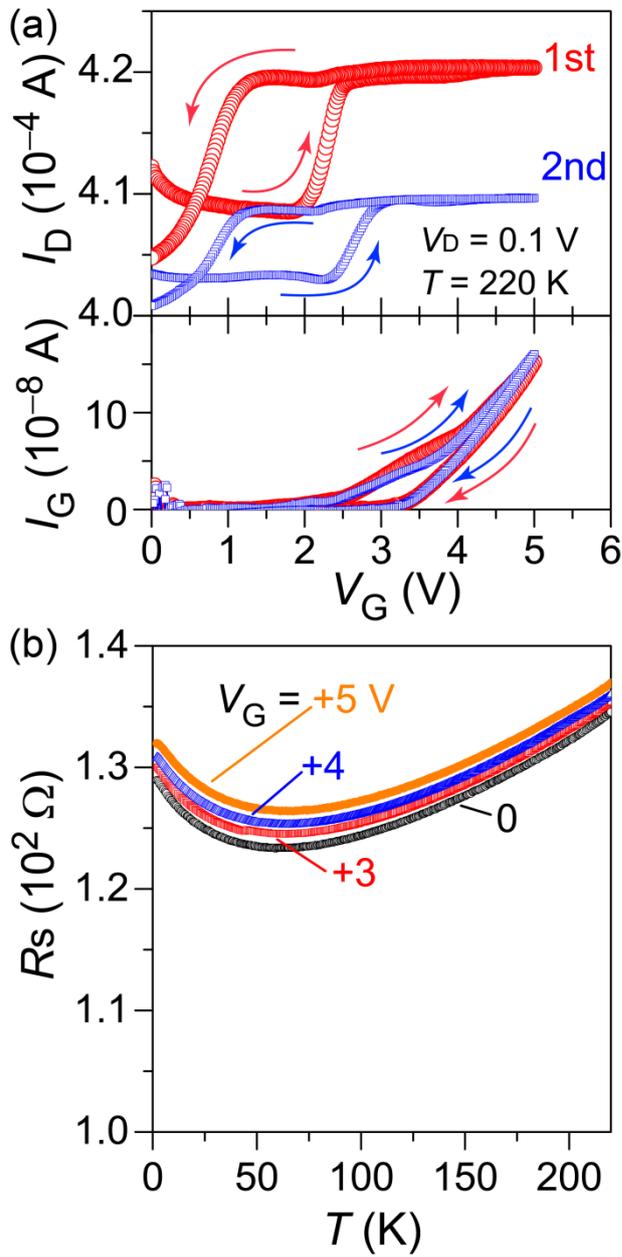

**Figure 7.** Electric transport properties of an EDLT with a *t*-FeS channel on CaF$_2$ under $V_G$. (a) Transfer curves of first (red, top) and second cycles (blue, top) and leakage currents (bottom) at 220 K. (b) Dependence of $R_s$ on $T$ under $V_G$ of 0 to +5 V.

**Table 1.** Structural parameters and chemical composition of $t$-FeS films. Values in parentheses for YSZ, CaF$_2$, and GaAs indicate the case of cube-on-cube heteroepitaxy.

| Substrate | $a_{sub}$ (Å) | Lattice mismatch $\Delta a$ (%) | [Fe]/[Se] | $a_{film}$ (Å) | $c_{film}$ (Å) | $\Delta\omega$ (deg.) | $R_{rms}$ (nm) |
|---|---|---|---|---|---|---|---|
| MgO | 4.213 | 14.36 | 1.2 | - | 5.070 | - | 1.8 |
| GaAs | 3.998 (5.654) | 7.95 (34.9) | 1.2 | - | 5.050 | - | 0.9 |
| STO | 3.905 | 7.61 | 1.1 | - | 5.070 | 4.35 | 1.2 |
| LSAT | 3.868 | 6.72 | 1.3 | 3.633 3.614 | 5.076 | 5.16 | 2.6 |
| CaF$_2$ | 3.862 (5.462) | 6.58 (32.6) | 0.9 | 3.668 | 5.092 | 2.00 | 1.6 |
| YSZ | 3.638 (5.139) | −1.16 (28.4) | 1.2 | - | 5.060 | 3.69 | 0.7 |

Supplementary Information for 'Stabilization and heteroepitaxial growth of metastable tetragonal FeS thin films by pulsed laser deposition'

Kota Hanzawa, Masato Sasase, Hidenori Hiramatsu, and Hideo Hosono

We would like to show a possible origin that can explain the unusual rotated domain by 37° observed in Fig. 4(c) in the following Figs. S1 and S2.

Figure S1 shows the in-plane atomic configuration of LSAT single-crystal substrate. As seen in the red domain (distance of 9.670 Å between an oxygen and (La,Al) atoms), we confirmed an angle of ~37° with respect to $a$/$b$ axis of LSAT, which is consistent with the result observed in Fig. 4(c). The distance of 9.670 Å roughly corresponds to ca. 2.5 times in-plane lattice of $t$-FeS ($a$ = 3.683 Å).

Figure S2 is the stacking model of $c$-axis oriented $t$-FeS on LSAT that we explained above. This is one of the possibilities that can explain the unusual rotational domain.

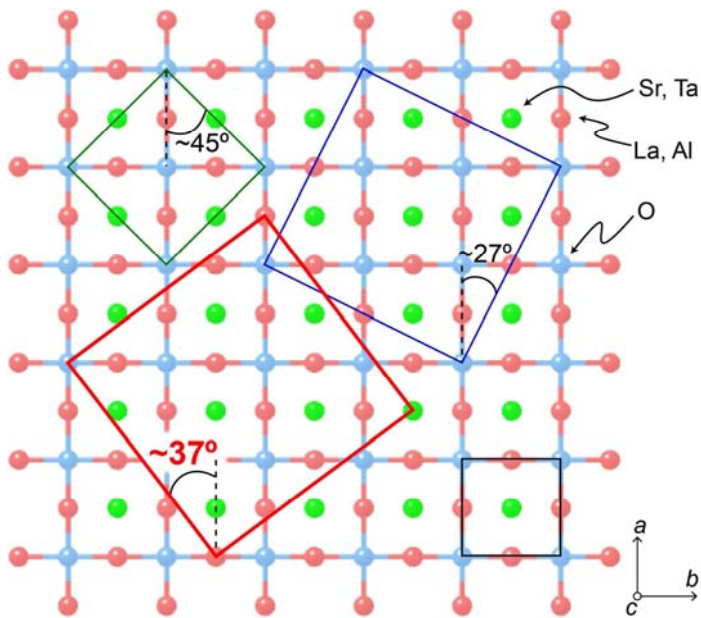

Figure S1. In-plane atomic configuration of LSAT single-crystal substrate. The black square denotes the half unit cell of LSAT. Three kinds of possible rotational domains are presented by red, blue, and green squares. One of them, indicated by red, is consistent with the result observed in Fig. 4(c).

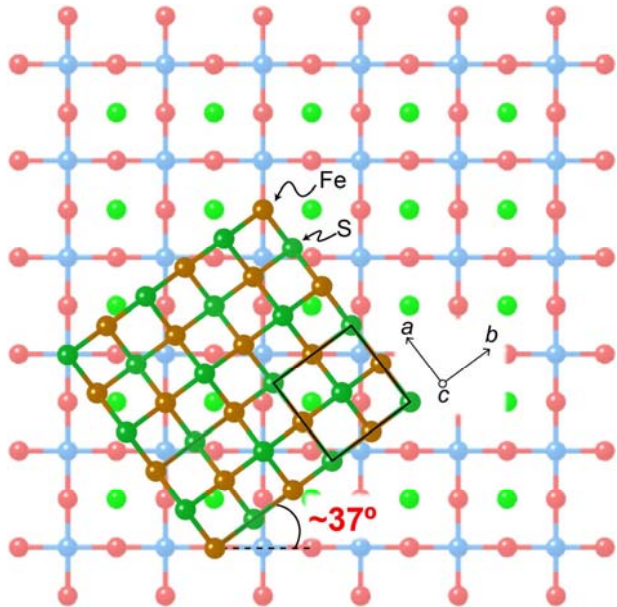

Figure S2. The stacking model of *t*-FeS layer on LSAT on the assumption of the red rotational domain in Fig. S1. The black square denotes the unit cell of *t*-FeS.